\documentclass[runningheads]{llncs}
\usepackage[T1]{fontenc}
\usepackage{graphicx}

\usepackage{xcolor, soul}
\usepackage{orcidlink}
\sethlcolor{green}

\usepackage[most]{tcolorbox}

\newtcolorbox{greybox}{
  colback=gray!10,
  colframe=gray!50,
  boxrule=0.4pt,
  arc=1mm,
  left=7pt,
  right=7pt,
  top=7pt,
  bottom=7pt
}

\usepackage[shortlabels]{enumitem}
\usepackage{hyperref}

\makeatletter
\renewcommand\footnotesize{%
   \@setfontsize\footnotesize{8pt}{10pt}%
   \abovedisplayskip 6\p@ \@plus2\p@ \@minus4\p@
   \abovedisplayshortskip \z@ \@plus\p@
   \belowdisplayshortskip 6\p@ \@plus\p@ \@minus3\p@
   \def\@listi{\leftmargin\leftmargini
               \topsep 3\p@ \@plus\p@\@minus\p@
               \parsep 2\p@ \@plus\p@\@minus\p@
               \itemsep \parsep}%
   \belowdisplayskip \abovedisplayskip
}
\makeatother

\begin{document}
%

\title{%
  Insights on Harmonic Tones from a Generative Music Experiment \\
  \ \\
  \small 15th International Workshop on Machine Learning and Music \\
   September 9, 2024, Vilnius, Lithuania}

%
%
\author{Emmanuel Deruty\inst{1,2} \and Maarten Grachten\inst{3}} 
%
\authorrunning{E. Deruty \and M. Grachten}
%
\institute{Sony Computer Science Laboratories - Paris, 6 rue Amyot, 75005 Paris, France \and
Department of Architecture, Design and Media Technology, Aalborg University, Rendsburggade 14, 9000 Aalborg, Denmark
\email{emmanuel.deruty@sony.com} \orcidlink{0009-0007-3912-5715}\and Contractor for Sony CSL Paris \orcidlink{0000-0002-9488-0840}}

\maketitle

\vspace{1cm}
\begin{greybox}
\begin{small}
Published version: Deruty, E., \& Grachten, M. (2026).
Insights on Harmonic Tones from a Generative Music Experiment.
In: Cerrato, M., Kalinauskait\.{e}, D.,
Luko\v{s}evi\v{c}ius, M., Pechenizkiy, M.,
\v{S}utien\.{e}, K. (eds),
\textit{Machine Learning and Principles and Practice of Knowledge Discovery in Databases}.
ECML PKDD 2024.
\textit{Communications in Computer and Information Science},
vol.~2559. Springer, Cham.
\url{https://doi.org/10.1007/978-3-032-25305-7_25}
\end{small}
\end{greybox}

\vspace{1cm}

\begin{abstract}

\normalsize

The ultimate purpose of generative music AI is music production. The studio-lab, a social form within the art-science branch of cross-disciplinarity, is a way to advance music production with AI music models. During a studio-lab experiment involving researchers, music producers, and an AI model for music generating bass-like audio, it was observed that the producers used the model's output to convey two or more pitches with a single harmonic complex tone, which in turn revealed that the model had learned to generate structured and coherent simultaneous melodic lines using monophonic sequences of harmonic complex tones. These findings prompt a reconsideration of the long-standing debate on whether humans can perceive harmonics as distinct pitches and highlight how generative AI can not only enhance musical creativity but also contribute to a deeper understanding of music.

\end{abstract}

\newpage

\section{Introduction}\label{sec:introduction}

The present contribution describes how analysis of the music stemming from a studio-lab experiment involving generative AI for music led to acknowledging the presence of simultaneous pitches in single harmonic complex tones in contemporary popular music, as well as the reexamination of the question: to what extent humans can `hear out' individual harmonics?

One outcome from the experiment was a 4-minute music piece titled Melatonin, described by Deruty and Grachten \cite{deruty2022melatonin} from a perspective of \textit{style}, citing, in particular, \textit{homophony within bass lines}, i.e., the presence of two or more melodic lines in monophonic tracks built from single harmonic complex tones. The present contribution elaborates on this aspect, providing more detailed observations of the Melatonin bass tracks, a historical context for the perception of simultaneous pitches in single harmonic complex tones, additional examples from contemporary popular music, and examples of research later stemming from the observations.

Section~\ref{sec:background} provides the background of the experiment, including the concept of the studio-lab, the participants, and the AI music generation model, BassNet. Section~\ref{sec:sessionresults} presents the observations derived from the experiment’s results. Section~\ref{sec:severalpitches} discusses these observations, offering historical context (Section~\ref{subsec:historicalcontext}) and examples from contemporary popular music (Section~\ref{subsec:incontpopmus}).
Supplementary material with audio and video examples is provided at the address: \begin{center}
\url{http://mml2024-suppl-mat.s3-website.eu-west-3.amazonaws.com/}
\end{center} 

\section{Background}\label{sec:background}

\subsection{The Studio-Lab}\label{subsec:studiolab}

Barry et al. \cite{barry2008logics} describe the studio-lab as a form of art-science collaboration that is transdisciplinary in nature. Transdisciplinarity involves practices that transcend disciplinary norms and focus on complex, real-world problem-solving \cite[p.~27]{barry2008logics}. Art-science, as a subset of cross-disciplinarity, is seen as an interdisciplinary space where the arts, sciences, and technologies intersect and overlap \cite[p.~39]{barry2008logics}. The studio-lab itself is characterized as a social space for hybrid experimental activity, where new media technologies are co-evolved with their creative applications \cite{born1995rationalizing}\cite[p.~2]{century1999pathways}. Key conditions for effective participation in the studio-lab include sustained collaboration, avoidance of accountability and service-subordination logic, fluid division of labor, and artist training for specialized facilities \cite[p.~39]{barry2008logics}. An early example of a music studio-lab is the 1949–1958 collaboration between engineer Pierre Schaeffer and musician Pierre Henry at the \textit{Club d'Essai} studio of the RTF~\cite{dhomont2001henry}.

\subsection{A CSL / Hyper Music Studio-Lab Experiment with AI}\label{subsec:studiolabAI}

The music research group of Sony Computer Science Laboratories (CSL) – Paris\footnote{\url{https://cslmusicteam.sony.fr/}} develops AI-based music technology to support music artists in their creative workflows \cite{deruty2022development}.

Hyper Music\footnote{\url{https://www.hyper-music.com/}} is a sound and music production company that produces music for ads, TV series, and feature films. The Hyper Music producers, Luc Leroy and Yann Mac\'e, define their style as electronic music with urban inflections. Hyper Music authored music for commercials commissioned by brands such as L'Or\'eal,  Adidas, Vichy, Honda, GoPro, and Chanel. Hyper Music cite globally successful artists such as Trent Reznor, Daft Punk, Kanye West, Skrillex, and Rosalia as influences.

At the end of 2021, CSL organized an art-science studio-lab experiment with Hyper Music, centered on BassNet~\cite{deruty2022melatonin} (Section~\ref{subsec:BassNet}). The session, which included two Hyper Music producers and two CSL researchers, adhered to the studio-lab criteria. CSL and Hyper Music had a long-standing collaboration dating back to 2018, with three participants having known each other since 1995. There was no financial interest involved, ensuring that the project was free from monetary constraints. Both CSL and Hyper Music collaborated on equal terms, avoiding any hierarchical division seen in previous projects \cite{deruty2022development}. The division of labor was fluid, allowing both parties to actively engage and intervene in each other's work. Moreover, Hyper Music had sufficient time to familiarize themselves with BassNet, enabling a seamless integration of their creative input. The session highlighted the \textit{co-evolution} of music and technology.


\subsection{BassNet}\label{subsec:BassNet}

The version of BassNet used in this study is a generative deep-learning model trained to learn the relationship between the bass track and the mix minus the bass in contemporary popular music \cite{grachten2020bassnet}. The model takes audio of the bass-less mix as input and predicts basslines as a pair of $f_0$ trajectories and Constant-Q Transform (CQT) log magnitude spectrograms as outputs. The $f_0$ and CQT spectrograms are combined to produce the bass track audio using additive synthesis of sine waves, where the harmonic series is determined by the predicted $f_0$, and the relative amplitude of the harmonics is inferred from the predicted CQT spectrum.

This version of BassNet is effectively a \textit{machine learning-powered synthesizer}, generating bass-style audio that harmonically and timbrally matches the input material. Like all synthesizers, BassNet includes several parameters. One such parameter is the \textit{Onset Threshold}, which allows the user to set the input level above which BassNet produces a new note, thereby controlling the density of output notes. Another parameter is \textit{Harmonics}, which enables the user to define the number of harmonics involved in the sonification. Additionally, there are three parameters for setting a low-pass filter, a classic component of synthesizers since the design of the Moog \cite{moog1965voltage}.

During the session, two additional parameters of the sonification process were implemented. One is called \textit{Harmonic Variation}, which was added to BassNet following discussions with the producers to achieve a more lively sound. This parameter controls the temperature of a (non-linear) softmax transformation of the CQT magnitudes to produce the relative strengths of the synthesized harmonics. It effectively enhances the amplitude variations of higher harmonics over time \cite[p.~8 and suppl. mat.]{deruty2022melatonin}. The other parameter sets the relative strength of odd and even harmonics. It was also implemented following discussions with the producers, who find this parameter useful whenever additive synthesis is involved.


\section{Session Outcome: Monophonic Bass Tracks with Multiple Pitches}\label{sec:sessionresults}

The song Melatonin is divided into two sections and features drum tracks, various tonal tracks, and eleven audio tracks from BassNet outputs, around which the song is centred. Of these eleven tracks, nine contain melodic patterns. The supplementary material includes interactive and video content detailing analyses of seven BassNet melodic tracks, the workflow, and the final result.

\begin{figure}[ht]
\includegraphics[width=\textwidth]{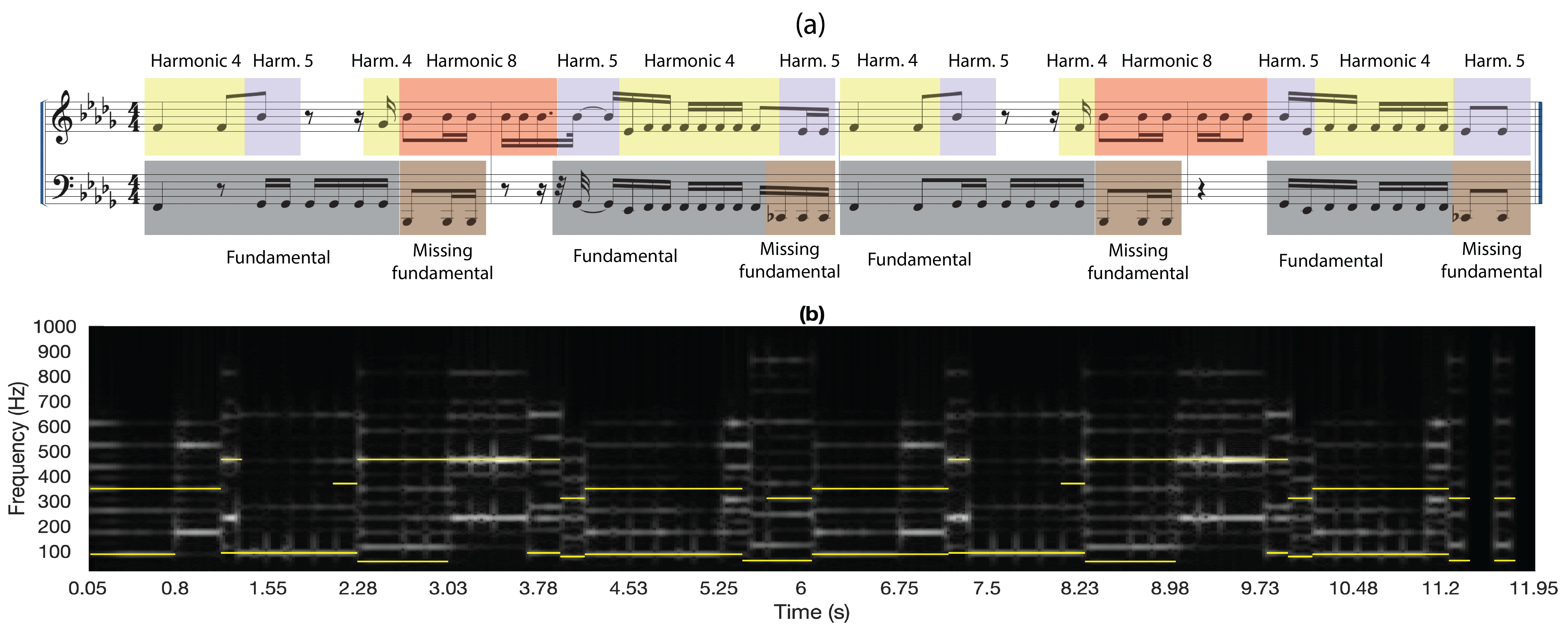}
\caption{Melatonin, section 2, bars 45--49, one BassNet output. (a) Perceptual transcription. Pitches that can be associated to different partials are highlighted using different colors. (b) STFT of the weighted audio. The horizontal yellow lines denote the pitch as transcribed in (a). Listen to this example: {\scriptsize \url{http://mml2024-suppl-mat.s3-website.eu-west-3.amazonaws.com/index.html\#LI_animation_12}}} \label{fig:Melatonin}
\end{figure}

Perceptual and signal analyses of the BassNet tracks were conducted post-session. Notably, simultaneous melodic lines were found in single BassNet outputs, an unexpected result given that BassNet's output for each frame is a single harmonic series, typically producing a single note and pitch. Analyses revealed that most BassNet outputs in Melatonin contain two or three simultaneous melodic lines, as illustrated in Figure~\ref{fig:Melatonin}. Further processing by the producers (not detailed in this paper) was applied to enhance the perception of multiple pitches.


In Figure~\ref{fig:Melatonin}, and the subsequent figures, spectral analyses are performed on weighted audio to account for human sensitivity variations across frequencies \cite{fletcher1933loudness}. Models such as ISO 226-2003 \cite{iso2262003} describe how tones are perceived as equally loud depending on frequency and sound pressure level. This study uses the ISO 226-2003 weighting to ensure that evaluated power spectra more accurately reflect the listener's perceived spectral profile.

In Figure~\ref{fig:Melatonin} as well as in the examples from the supplementary material, the following aspects of the signal may account for the presence of simultaneous coherent melodic lines:

\begin{enumerate}[label={(\arabic*)}]

    \item Harmonics loud enough to be individually audible.
    \item Use of only odd harmonics, combined with a weak fundamental.
    \item Inharmonicity (in one example only).

\end{enumerate}

Aspect (1) arises from both the sonification process and the model's behaviour. While the weakness of the first two harmonics in aspect (2) enhances the audibility of upper harmonics, the relative strengths of these harmonics are determined by the model's generated CQTs. In videos 1 and 2 from the supplementary material, the top harmonic line consistently follows harmonic 3, whereas the remaining videos (including the example in Figure~\ref{fig:Melatonin}) show how the model selectively favours changing harmonics to produce additional coherent melodic lines.
This behavior is enhanced by the Harmonic Variation dial.

Aspect (2) contributes to the perception of multiple pitches in videos~4 to~7. Here, the lowest strong partial is an odd harmonic of the original $f_0$, with harmonic spacing of $2f_0$. This setup favours the perception of two pitches \cite{yost2009pitch}.

Aspect (3) applies to only one example, complicating transcription and analysis. As noted in Bregman (1996) \cite[p.~41]{bregman1996demonstrations}, when partials deviate from harmonic positions, both the fundamental and `some other partials than the fundamental' dominate \cite{jarvelainen2000effect}.

A key point is that sequences of harmonics contain independent horizontal structures originating from the model. While the missing even harmonics and fundamental, as well as the Harmonic Variation dial emphasise the phenomenon, the source of the structures is the predicted CQT spectrogram.


\newpage

\section{Several Pitches in Harmonic Complex Tones}\label{sec:severalpitches}

The presence of simultaneous melodic lines in sequences of unique complex tones, stemming from the perception of several pitches in single harmonic complex tones, links to a discussion that may be traced back to the 17\textsuperscript{th} century.

\subsection{Historical Context}\label{subsec:historicalcontext}

The question of whether humans can `hear out' individual harmonics has been long debated. In 1636, Mersenne first suggested that listeners might discern the multiple frequencies produced by vibrating strings \cite{mersenne1636harmonie}, influencing Rameau's theory of harmony \cite[pp.~13--14]{rameau1750demonstration}. This idea became central to the 19\textsuperscript{th}-century Ohm-Seebeck dispute, where Ohm argued that only the fundamental partial contributes to a tone's perceived pitch \cite{turner1977ohm}. Helmholtz supported this view, suggesting upper partials primarily affect timbre \cite[pp.~58--59]{helmoltz1885sensations}. Experiments indicate humans can hear only the first 5-8 harmonics, though their audibility is contested \cite{moore2012introduction}. Researchers like Plomp argue that harmonics can be discerned under favorable conditions, challenging Helmholtz's stance \cite{plomp1964ear}. Dixon Ward criticizes the frequency analyzer hypothesis as a `quarter-truth', applicable only under limited conditions \cite{dixonward1970}.

\subsection{In Contemporary Popular Music}\label{subsec:incontpopmus}

\begin{figure}[ht]
\includegraphics[width=\textwidth]{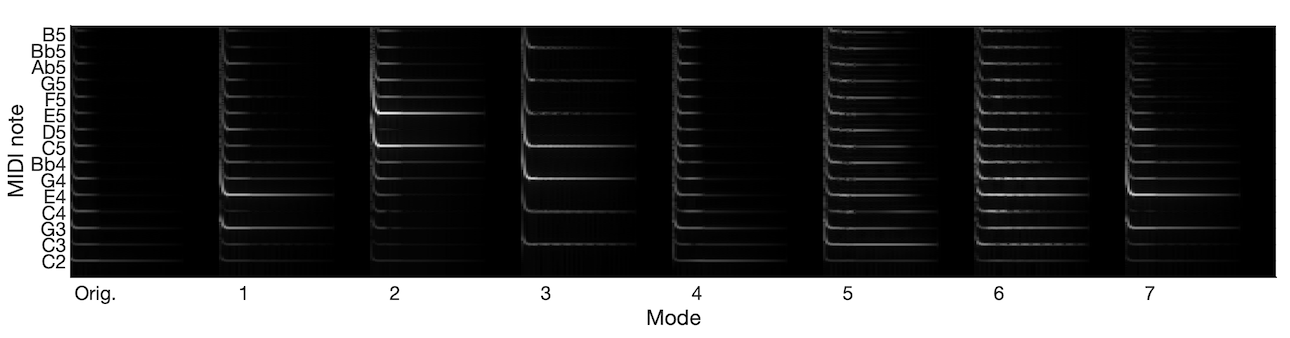}
\caption{STFT for the seven `modes' of the `808 Woofer Warfare' patch from the Seismic Shock library in Omnisphere, weighted audio. {\scriptsize \url{http://mml2024-suppl-mat.s3-website.eu-west-3.amazonaws.com/index.html\#seismic_shock}}
} \label{fig:Warfare}
\end{figure}

In studio music production, many technologies make obtaining sounds with loud upper harmonics straightforward: equalisers, oscillator modulation in synthesisers, ring modulation, and more. As illustrated in Figure~\ref{fig:Warfare}, the `808 Woofer Warfare' patch from the Seismic Shock\footnote{\url{https://www.sonicextensions.com/seismic-shock/}} Omnisphere\footnote{\url{https://www.spectrasonics.net/products/omnisphere/}} library, a Roland 808\footnote{\url{https://www.roland.com/us/promos/roland_tr-808/}}-style bass generator, even offers presets (`modes') in which particular upper harmonics are selectively amplified.

\begin{figure}[ht]
\includegraphics[width=\textwidth]{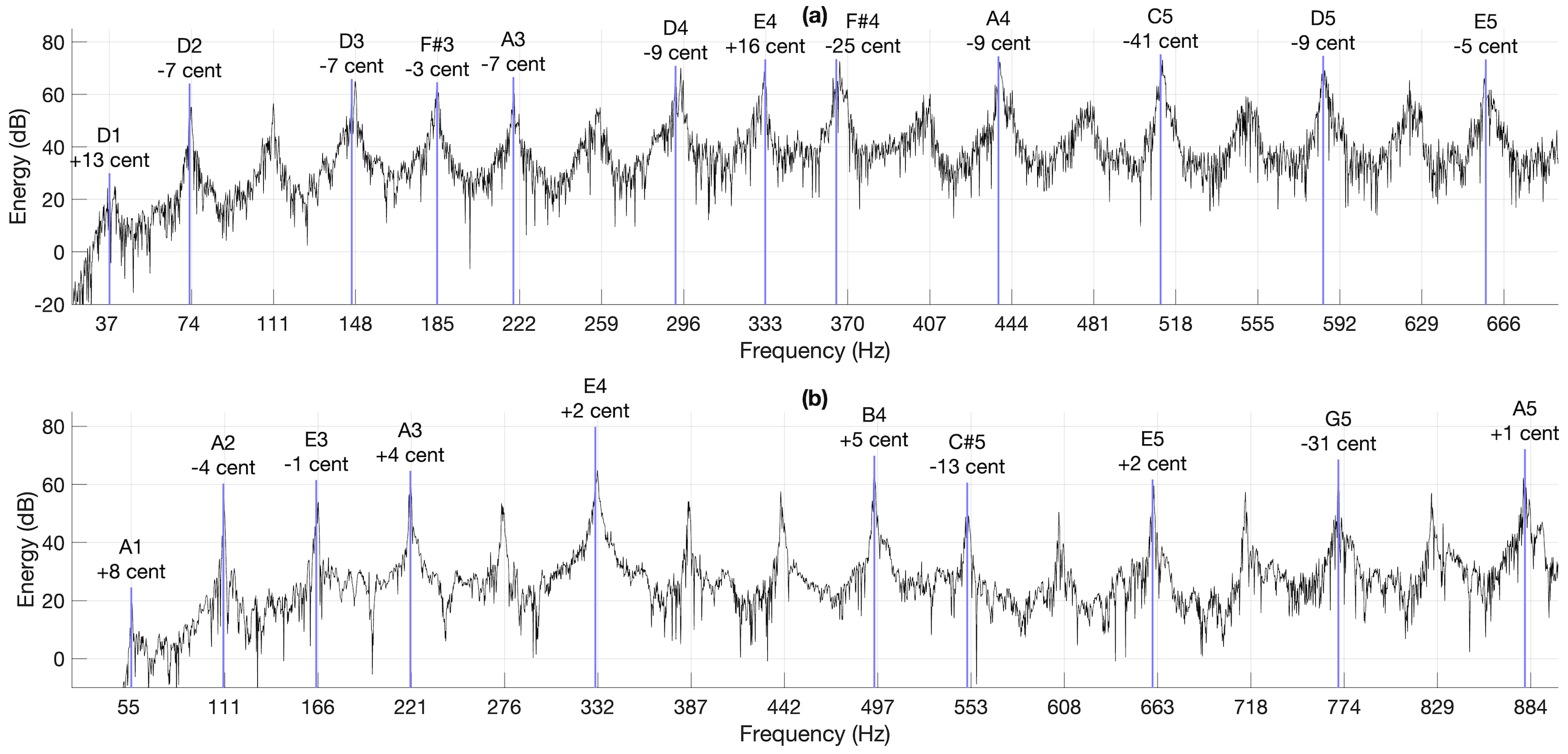}
\caption{(a) Alt-J, `Hunger of the Pine', 4'30 to 4'47, power spectrum of the weighted audio. (b) Guitar power chord, power spectrum of the weighted audio.} \label{fig:Hunger}
\end{figure}

Figure~\ref{fig:Hunger} shows two examples of harmonic complex tones where simultaneous pitch values can be heard. The first example is the end part of a song from British band Alt-J. A listening test involving musicians as well as non-musicians asked respondents how many simultaneous pitches they could hear in this excerpt. 6/20 could hear one pitch, 4/20 could hear two pitches, 6/20 could hear two pitches, and 4/20 could hear four pitches, leading to a mean of 2.22 perceived pitches. Other listening tests show that the mean number of perceived pitches in tones used by Hyper Music in their independent activity ranges from 1.7 to 2.45. 

\newpage

The second example is a guitar power chord generated from an Epiphone Emily the Strange guitar sample,\footnote{\url{https://www.karoryfer.com/karoryfer-samples/emilyguitar}} processed with  Native Instrument's Guitar Rig Rammfire amp emulation unit.\footnote{\url{https://www.native-instruments.com/en/products/komplete/guitar/guitar-rig-7-pro/amps-and-cabinets/}} Although it is perceived as a chord, it is a single harmonic complex tone.

\newpage

In commercial contemporary popular music, bass harmonics may also form independent lines. For example, \cite[suppl. mat.]{deruty2022melatonin} highlights two contemporary popular music tracks, The Cure's `The Holy Hour' and Judas Priest's `Bloodstone', where a second independent melodic line can be heard based on audible upper harmonics. In the first example, the upper harmonics are emphasized by the use of a flanger. In the second example, they are made audible by a combination of EQ and hard plectrum strokes.





\section{Conclusion}\label{sec:conclusion}


This contribution demonstrates how a studio-lab experiment led to the unexpected result of the machine learning-powered BassNet synthesizer generating two or more independent melodies from a sequence of individual harmonic complex tones. While the ability to `hear out' individual harmonics has traditionally been considered valid only under specific conditions, the experiment suggests that these conditions may be less restrictive in contemporary popular music. In fact, the Hyper Music producers themselves consider multi-pitch structured timbres essential to their sound.
Changes to the BassNet synthesis method made in response to their requirement for rich timbres, brought about not only multi-pitch structured timbres in BassNet outputs, but also revealed that the upper harmonics in the generated content can contribute to independent melodic lines.


Following the experiment, further research has been conducted on the matter. Planned publications include studies exploring how Hyper Music producers and electronic musician Vitalic use mainstream technology to create complex tones that lead to the perception of simultaneous pitch values. 

More generally, the Hyper Music/CSL studio-lab experiment suggests that generative AI can contribute to a better understanding of music under the right conditions. The findings confirm that `collaborative projects between artists and scientists may provoke and enrich scientific research, triggering unforeseen directions' \cite[p.~39]{barry2008logics}.




\begin{credits}
\subsubsection{\ackname} The authors would like to thank Luc Leroy and Yann Mac\'e from Hyper Music for their contribution and helpful insights.


\subsubsection{\discintname}
The authors have no competing interests to declare that are relevant to the content of this article.
\end{credits}

%
%
%
\newpage

\bibliographystyle{splncs04}
\bibliography{MML_Vilnius}

@article{barry2008logics,
  title={Logics of interdisciplinarity},
  author={Barry, Andrew and Born, Georgina and Weszkalnys, Gisa},
  journal={Economy and society},
  volume={37},
  number={1},
  pages={20--49},
  year={2008},
  publisher={Taylor \& Francis}
}

@book{born1995rationalizing,
  title={Rationalizing culture: IRCAM, Boulez, and the institutionalization of the musical avant-garde},
  author={Born, Georgina},
  year={1995},
  publisher={Univ of California Press}
}

@book{bregman1996demonstrations,
  title={Demonstrations to Accompany Bregman's Auditory Scene Analysis},
  author={Bregman, Albert S. and Ahad, Pierre A.},
  year={1996},
  publisher={Cambridge, MA, and London: MIT Press}
}

@book{century1999pathways,
  title={Pathways to innovation in digital culture},
  author={Century, Michael},
  year={1999},
  publisher={McGill University, Montreal}
}

@article{deruty2022development,
  title={On the development and practice of {AI} technology for contemporary popular music production},
  author={Deruty, Emmanuel and Grachten, Maarten and Lattner, Stefan and Nistal, Javier and Aouameur, Cyran},
  journal={TISMIR},
  volume={5},
  number={1},
  year={2022},
  publisher={Ubiquity Press}
}

@inproceedings{deruty2022melatonin,
  author       = {Deruty, Emmanuel and
                  Grachten, Maarten},
  title        = {``{M}elatonin'': A Case Study on {AI}-induced Musical 
                   Style},
  booktitle    = {{Proceedings of the 3rd Conference on AI Music 
                   Creativity}},
  year         = 2022,
  publisher    = {AIMC},
  month        = sep}

@incollection{dhomont2001henry,
  title={Henry, {P}ierre},
  author={Dhomont, Francis},
  booktitle={Grove Music Online},
  year={2001}
}

@incollection{dixonward1970,
  author    = "Dixon Ward, W.",
  title     = "Musical Perception",
  booktitle = "Foundations of Modern Auditory Theory",
  publisher = "Academic Press",
  year      = "1970",
  volume    = "1",
  pages     = "405-446",
  editor = "Tobias, Jerry"
}

@article{fletcher1933loudness,
	author = {Fletcher, Harvey and Munson, Wilden A},
	journal = {Bell System Technical Journal},
	number = {4},
	pages = {377--430},
	publisher = {Wiley Online Library},
	title = {Loudness, its definition, measurement and calculation},
	url = {https://ieeexplore.ieee.org/document/6771028},
	volume = {12},
	year = {1933}
}

@article{grachten2020bassnet,
  title={BassNet: A Variational Gated Autoencoder for Conditional Generation of Bass Guitar Tracks with Learned Interactive Control},
  author={Grachten, Maarten and Lattner, Stefan and Deruty, Emmanuel},
  journal={Applied Sciences},
  volume={10},
  number={18},
  pages={6627},
  year={2020},
  publisher={Multidisciplinary Digital Publishing Institute}
}

@book{helmoltz1885sensations,
  title={On the sensations of tone as a physiological basis for the theory of music},
  author={Helmholtz, Hermann Ludwig Ferdinand von},
  publisher={Longmans, Green, and Co.},
  year={1885}
}

@techreport{iso2262003,
    author = {ISO},
    type = {Standard},
    key = {ISO 226:2003},
    year = {2003},
    title = {{Normal equal-loudness level contours-ISO 226: 2003}},
    address = {Geneva, Switzerland},
    institution = {International Organization for Standardization}
}

@inproceedings{jarvelainen2000effect,
  title={The effect of inharmonicity on pitch in string instrument sounds.},
  author={J{\"a}rvel{\"a}inen, Hanna and Verma, Tony S and V{\"a}lim{\"a}ki, Vesa},
  booktitle={Proceedings of the 26th International Computer Music Conference, ICMC 2000, Berlin, Germany},
  year={2000}
}

@book{mersenne1636harmonie,
    title={Harmonie Universelle},
    author={Mersenne, Marin},
    year={1636},
    publisher={S\'ebastien Cramoisy, Pierre Ballard et Richard Charlemagne}
}

@inproceedings{moog1965voltage,
  title={A voltage-controlled low-pass high-pass filter for audio signal processing},
  author={Moog, Robert A},
  booktitle={Audio Engineering Society Convention 17},
  year={1965},
  organization={Audio Engineering Society}
}

@book{moore2012introduction,
  title={An introduction to the psychology of hearing},
  author={Moore, Brian CJ},
  year={2012},
  publisher={Emerald Group Publishing Limited}
}

@article{plomp1964ear,
  title={The ear as a frequency analyzer},
  author={Plomp, Reinier},
  journal={The Journal of the Acoustical Society of America},
  volume={36},
  number={9},
  pages={1628--1636},
  year={1964},
  publisher={Acoustical Society of America}
}

@book{rameau1750demonstration,
  title={D\'emonstration du principe de l'harmonie, servant de base \`a tout l'art musical th\'eorique et pratique},
  author={Rameau, Jean Philippe},
  year={1750},
  publisher={Paris: Durand et Pissot}
}

@article{turner1977ohm,
  title={The {Ohm-Seebeck dispute}, {Hermann von Helmholtz}, and the origins of physiological acoustics},
  author={Turner, R Steven},
  journal={The British Journal for the History of Science},
  volume={10},
  number={1},
  pages={1--24},
  year={1977},
  publisher={Cambridge University Press}
}

@article{yost2009pitch,
  title={Pitch perception},
  author={Yost, William A},
  journal={Attention, Perception, \& Psychophysics},
  volume={71},
  number={8},
  pages={1701--1715},
  year={2009},
  publisher={Springer}
}

\end{document}